\documentclass[12pt]{article}

\usepackage{amsmath}
\usepackage{cite}

\title{Frame-like gauge invariant Lagrangian formulation of massive
fermionic higher spin fields in $AdS_3$ space}

\author{I.L. Buchbinder${}^{ab}$\thanks{joseph@tspu.edu.ru},
T.V. Snegirev${}^a$\thanks{snegirev@tspu.edu.ru}, Yu.M.
Zinoviev$^c$\thanks{Yurii.Zinoviev@ihep.ru}
\\[0.5cm]
\it ${}^a$Department of Theoretical Physics,\\
\it Tomsk State Pedagogical University,\\
\it Tomsk 634061, Russia\\[0.3cm]
\it ${}^b$National Research Tomsk State University, Russia\\[0.3cm]
\it ${}^c$Institute for High Energy Physics,\\
\it Protvino, Moscow Region, 142280, Russia}

\date{}

\begin{document}

\maketitle

\begin{abstract}
We construct the frame-like gauge-invariant Lagrangian formulation
for massive fermionic arbitrary spin fields in three-dimensional $AdS$
space. The Lagrangian and complete set of gauge transformations are
obtained. We also develop the formalism of gauge-invariant curvatures
for the massive theory under consideration and show how the Lagrangian
is formulated in their terms. The massive spin-5/2 field is discussed
as an example.

\end{abstract}

\thispagestyle{empty}
\newpage
\setcounter{page}{1}

\newpage
\section*{Introduction}

Despite the significant progress in the theory of massless
higher-spin fields in various dimensions (see reviews
\cite{Vasiliev96.11,BekCnockIazeolVasil05.03,Vasiliev04.01}, see
also \cite{DidenSkvorts14.01} for details), the construction of full
higher spin field theory is still far from being complete. In
particular it remains unclear how the non-linear theory of massive
higher spins should look like. Investigations of massive higher spin
interactions is very important in-particular for understanding the
relation between higher spins and (super)string theories that is
assumed to realize a kind of spontaneous symmetry breaking
mechanism. Thus a natural framework for such investigations is a
gauge invariant formulation for massive higher spin fields (similar
to the one appearing in string field theory) that becomes possible
due to introduction of appropriate number of Stueckelberg fields.
Currently gauge invariant description for the free massive
higher-spin fields is rather well studied, moreover various authors
have developed different ways of such description
\cite{Metsaev06.09,Zinoviev08.08,PonomVasil10.01,BuchKrykh05.05,
BuchKrykhRyskTak06.03}.

At the same time studying the massive higher spin interactions in
arbitrary dimensions appears to be even more complicated that the
ones for the massless fields. Therefore it would be instructive to
investigate a structure of the massive higher spin theories coupled
to external fields or among themselves in three-dimensional space
where situation becomes simpler then in a space of arbitrary
dimension. As a result we can get a nice playground to gain useful
experience and possibilities for generalizations. In particular it
turns out that in $d=3$ there exist examples of interacting models
with finite number of higher spin fields
\cite{AragDeser84,Blencowe89,CamFredPfenTheis10.08,ChenLongWang12.09}.
Moreover the specific properties of three dimensional space allow us
to construct a more exotic higher spin models
\cite{BergshHolmTowns09.11}. Therefore we can  expect that the
massive higher-spin theory also become easier in $d=3$.

First Lagrangian formulation for massive higher-spin fields in $d=3$
was considered in \cite{TyutVasil97.04} however in gauge invariant
form it has been developed for bosonic fields only
\cite{Zinoviev12.05,BuchSnegZin12.07,BuchSnegZin12.08}. In this work
we fill this gap and give gauge invariant formulation for massive
fermionic higher spins in three-dimensional $AdS_3$ space.
In this we think that the most convenient formalism is the frame-like
one \cite{Vasiliev87,LV88,Vas88}, which in particular allows one to
work with an explicitly invariant objects.

The paper is organized as follows. In section 1 we collect the basic
information about frame-like formulation of free massless higher
spins in $AdS_3$ focusing on the fermionic fields only. In section 2
we give the frame-like gauge-invariant formulation for free massive
fermionic higher spins in $AdS_3$. Here at first we discuss a
field content  which we need to have gauge invariant description
of massive fields. Then in terms of these fields we derive the
Lagrangian and corresponding gauge transformations. We also
construct the full set of gauge invariant linearized curvatures
and show how the free Lagrangian can be rewritten in their terms.
At the end of section 2 we consider the concrete example of massive
spin-5/2 field. In conclusion we summarize the main points.

\section{Massless fermionic fields}

In this section we briefly review a frame-like formulation of
massless higher spin fermionic fields in $AdS_3$ space. The
Lagrangian, gauge transformation and gauge invariant curvatures for
free fields are given.

Frame-like formulation of massless higher-spin fields can be treated
as generalization of the frame formulation of gravity in terms of
vielbein field $h_\mu{}^a$ and Lorentz connection
$\omega_\mu{}^{a,b}$. Such generalization was successfully developed
both for the bosonic fields and for the fermionic ones
\cite{Vasiliev87,LV88,Vas88}. We will focus here only on half-integer
spins, the details for integer spins can be found in
\cite{Vasiliev87,LV88}.

It is known \cite{Vasiliev87} that in four dimensions the field
with half-integer spin $s > 3/2$ in the frame-like formulation is
described by a total bunch of spin-tensor one-forms (here we omit
spinor
index)
$$
\Psi^{a_1...a_{s-3/2},b_1...b_k} = dx^\mu
\Psi_\mu{}^{a_1...a_{s-3/2},b_1...b_k}, \qquad 0 \leq k \leq s-3/2
$$
These fields on Latin indices form an irreducible representations
of the Lorentz group, i.e. satisfy the symmetry of two-row Young
tableaux and are $\gamma$-traceless. Note that for $k=0$ we have the
generalized vielbein field $\Psi^{a_1...a_{s-3/2}}$ while for other
values of $k$ we have so-called extra fields. An important
property of $d=3$ case is the absence of these extra fields that
greatly simplifies the calculations in construction of consistent
higher spin field theories. Moreover the description can be simplified
even more by the use of multispinor frame-like formalism where all
fields are still one-forms but with all local indices replaced by the
spinor ones \cite{Blencowe89}. The main object then is completely
symmetric multispinor one-form $\Psi^{\alpha(n)}$ (see Appendix for
notations and conventions) with $n=2(s-1)$ (where  $s$ is
half-integer and $n$ is odd) which is equivalent to the completely
symmetric spin-tensor $\Psi^{a_1...a_{s-3/2}}$ satisfying
$\gamma$-tracelessness condition $(\gamma \Psi)^{a_1 \dots a_{s-5/2}}
= 0$.

Free Lagrangian being three-form in three dimensional $AdS_3$ space
looks as follows \cite{Vasiliev96.11,Blencowe89}
\begin{equation}\label{MLLagr}
{\cal{L}}_0 = i\kappa_n [ \Psi_{\alpha(n)} D \Psi^{\alpha(n)}
+ \frac{n\lambda}{2}  \Psi_{\alpha(n-1)\beta} e^\beta{}_\gamma
\Psi^{\alpha(n-1)\gamma}]
\end{equation}
here $D=dx^\mu D_\mu$ is $AdS_3$ covariant derivative, one-form
$e^{\alpha(2)}$ is $AdS_3$ background vielbein (see Appendix for
details) and $\kappa_n = (-1)^{\frac{n+1}{2}}$. The Lagrangian
(\ref{MLLagr}) is invariant under the gauge transformations
$$
\delta_0 \Psi^{\alpha(n)} = D \xi^{\alpha(n)} + \frac{\lambda}{2}
e^\alpha{}_\beta \xi^{\alpha(n-1)\beta}
$$
with zero-form gauge parameter $\xi^{\alpha(n)}$. For this field
$\Psi^{\alpha(n)}$ there exists the two-form gauge invariant object
(curvature)
$$
{\cal{R}}^{\alpha(n)} = D \Psi^{\alpha(n)} + \frac{\lambda}{2}
e^\alpha{}_\beta \Psi^{\alpha(n-1)\beta}
$$
which satisfies the Bianchi identity
$$
D{\cal{R}}^{\alpha(n)} = - \frac{\lambda}{2} e^\alpha{}_\beta
{\cal{R}}^{\alpha(n-1)\beta}
$$
Using this curvature the free Lagrangian can be rewritten as follows
$$
{\cal{L}}_0 = i\kappa_n \Psi_{\alpha(n)} {\cal{R}}^{\alpha(n)}
$$
In order to verify gauge invariance of the Lagrangian in such form we
should use the Bianchi identity. It is easy to see that the
curvature just gives the equations of motion for the Lagrangian
(\ref{MLLagr}). Note that the possibility to work in terms of the
curvatures is a peculiarity of the frame-like formalism.

\section{Massive fermionic fields}

In this section we develop the frame-like gauge invariant
description for massive arbitrary half-integer spin $s\geq3/2$ in
$AdS_3$ space. To provide the gauge invariance we introduce the
auxiliary Stueckelberg fields and construct the Lagrangian in their
terms. Then we generalize the formalism of gauge-invariant
curvatures for massive fields. General formulation is illustrated on
the example of massive spin-5/2 theory.

\subsection{Lagrangian formulation}

Gauge invariant Lagrangian formulation of massive fields is based on
introduction of the auxiliary Stueckelberg fields. We follow the
procedure proposed in \cite{Zinoviev08.08} to use the minimal number
of such fields. In the case under consideration the full set of field
variables include the following one-forms
$\Psi^{\alpha(n)},\;n=1,3,...,2(s-1)$ (each one with its own gauge
transformations) and zero-form $\phi^\alpha$. We will look for the
free Lagrangian for massive field as the sum of kinetic terms for all
these fields as well as the most general mass-like terms gluing them
together:
\begin{eqnarray}\label{MVLagr}
{\cal{L}}_0 &=& i \sum_{n=1}^{2(s-1)} \frac{\kappa_n}{2}
\Psi_{\alpha(n)} D \Psi^{\alpha(n)} + \frac{i}{2} \phi_\alpha
E_2{}^\alpha{}_\beta D \phi^\beta \nonumber \\
&& + i \sum_{n=3}^{2(s-1)} \kappa_n a_n \Psi_{\alpha(n)}
e^{\alpha\alpha} \Psi^{\alpha(n-2)} + ia_0 \Psi_{\alpha(1)}
E_2{}^\alpha{}_\beta \phi^\beta \nonumber \\
&& + i \sum_{n=1}^{2(s-1)} \frac{\kappa_n b_n}{2} \Psi_{\alpha(n)}
e^\alpha{}_\beta \Psi^{\alpha(n-1)\beta} + i \frac{b_0}{2} \phi_\alpha
E_3\phi^\alpha
\end{eqnarray}
here $\kappa_n = (-1)^{\frac{n+1}{2}}$ and $a_n,b_n$ are free
parameters to be determined. The most general form of the
corresponding gauge transformations look like
\begin{eqnarray}\label{GT}
\delta_0 \Psi^{\alpha(n)} &=& D \xi^{\alpha(n)} + \alpha_n
e^\alpha{}_\beta \xi^{\alpha(n-1)\beta} + \beta_n e^{\alpha\alpha}
\xi^{\alpha(n-2)} + \gamma_n e_{\beta\beta} \xi^{\alpha(n)\beta\beta}
\nonumber \\
\delta_0 \phi^\alpha &=& \alpha_0 \xi^\alpha
\end{eqnarray}
where $\alpha_n,\beta_n,\gamma_n$ are also free parameters. Our aim
is to find the restrictions on the parameters in the Lagrangian and
gauge transformations providing the gauge invariance of the Lagrangian
(\ref{MVLagr}) under the transformations (\ref{GT}).

First of all we consider variations of the Lagrangian that are of the
first order in derivatives. This allows us to express the parameters
$a_n,b_n$ in the Lagrangian through parameters in the gauge
transformations
\begin{equation}\label{LagrParam}
a_n = \frac{n(n-1)}{2} \beta_n, \qquad b_n = n \alpha_n,
\qquad a_0 = \alpha_0
\end{equation}
and also imposes one restriction on the parameters of the gauge
transformations
\begin{equation}\label{GTParamBeta}
\beta_n = \frac{2}{n(n-1)} \gamma_{n-2}
\end{equation}
Note that this condition is valid because $\beta_n$ is defined for
$3\leq n\leq 2(s-1)$ and $\gamma_n$ for $1\leq n\leq 2(s-2)$. The
remaining free parameters $\alpha_n,\gamma_n$ are fixed from the
invariance conditions under the transformations (\ref{GT}) for the
variations without derivatives. Direct calculations yield the
following equations
\begin{align}\label{Cond}
& 2(n+2)a_n\alpha_n - 2b_{n-2}\gamma_{n-2} = 0 \qquad n \geq3
\nonumber \\
& 2(n-2)a_n\alpha_{n-2} - (n-1)(n+2)b_n\beta_n = 0 \qquad n\geq3
\nonumber \\
& - n\lambda^2 - 4a_n\gamma_{n-2} +
4b_n\alpha_n + 2n(n+3)a_{n+2}\beta_{n+2} = 0 \qquad n\geq3 \\
& - \lambda^2 + 4b_1\alpha_1 + 8a_3\beta_3 + a_0\alpha_0 = 0
\qquad n=1 \nonumber \\
& 3a_0\alpha_1 + b_0\alpha_0 = 0 \nonumber
\end{align}
Last equation allows us to express the only remaining free parameter
in the Lagrangian $b_0$
$$
b_0 = - 3\alpha_1
$$
Taking into account the relations
(\ref{LagrParam}),(\ref{GTParamBeta}) we see that first two
equations in (\ref{Cond}) are identical and lead to a simple
recurrent relation
$$
(n-2)\alpha_{n-2} - (n+2)\alpha_n = 0
$$
Denoting the maximal value of $n=2(s-1)=\hat{s}$ and expressing
everything through $\alpha_{\hat{s}}$ we get the general solution
for ${\alpha}_n$ in the form
\begin{equation}\label{GTParamAlpha}
\alpha_n = \frac{\hat{s}(\hat{s}+2)}{n(n+2)} \alpha_{\hat{s}}
\end{equation}
Now let us consider the third equation (\ref{Cond}) containing two
parameters
\begin{equation}\label{Cond3}
- n\lambda^2 + 4n\alpha_n{}^2 - 4 \gamma_{n-2}{}^2 +
\frac{4n(n+3)}{(n+1)(n+2)} \gamma_n{}^2 = 0
\end{equation}
Here we have used the relations
(\ref{LagrParam}),(\ref{GTParamBeta}). First, note that the
$\gamma_n$ is absent for $n=\hat{s}$. Therefore one can express
$\alpha_{\hat{s}}$ through $\gamma_{\hat{s}-2}$ (as we will see later
the parameter $\gamma_{\hat{s}-2}$ will remain as the only free one
and will play the role of a mass parameter)
\begin{equation}\label{GTParamAlphaS}
\alpha_{\hat{s}}{}^2 = \frac{1}{\hat{s}} \gamma_{\hat{s}-2}{}^2 +
\frac14\lambda^2
\end{equation}
Taking into account (\ref{GTParamAlpha}), we see that the relation
(\ref{Cond3}) is a recurrent equation for parameters $\gamma_n$.
General solution looks like
\begin{equation}\label{GTParamGamma}
\gamma_{n-2}^2 = \frac{(\hat{s}-n+2)(\hat{s}+n+2)}{4n(n+1)}
\big[ m^2 + \frac14 (\hat{s}-n)(\hat{s}+n)\lambda^2 \big]
\end{equation}
where we introduced mass parameter $m^2 = \hat{s} \gamma_{s-2}{}^2$.
Finally from the fourth equation of system (\ref{Cond}) it follows
that
\begin{equation}\label{GTParamAlpha0}
\alpha_0{}^2 = \frac{(\hat{s}+1)(\hat{s}+3)}{2} \big[ m^2 + \frac14
(\hat{s}-1)(\hat{s}+1)\lambda^2 \big]
\end{equation}
Let us also introduce convenient combination
\begin{equation}\label{paraM}
M^2 = m^2 + \frac14 \hat{s}^2 \lambda^2
\end{equation}
so that now
\begin{equation}\label{newAlpha}
\alpha_n = \frac{(\hat{s}+2)}{n(n+2)} M
\end{equation}
Thus all parameters are found and are defined by
(\ref{LagrParam}),(\ref{GTParamBeta}),
(\ref{GTParamGamma})-(\ref{newAlpha}) with $m^2$
as the only free one. Then the final expression for the Lagrangian
(\ref{MVLagr}) takes the form:
\begin{eqnarray}\label{MVLagrNorm}
{\cal{L}}_0 &=& i \sum_{n=1}^{2(s-1)} \frac{\kappa_n}{2}
\Psi_{\alpha(n)} D \Psi^{\alpha(n)} + \frac{i}{2} \phi_\alpha
E_2{}^\alpha{}_\beta D \phi^\beta \nonumber \\
&& + i \sum_{n=3}^{2(s-1)} \kappa_n m_n \Psi_{\alpha(n)}
e^{\alpha\alpha} \Psi^{\alpha(n-2)} + im_0 \Psi_{\alpha(1)}
E_2{}^\alpha{}_\beta \phi^\beta \nonumber \\
&& + i\sum_{n=1}^{2(s-1)}
\frac{(\hat{s}+2)\kappa_nM}{2(n+2)} \Psi_{\alpha(n)}
e^\alpha{}_\beta \Psi^{\alpha(n-1)\beta} - i
\frac{(\hat{s}+2)M}{2} \phi_\alpha E_3 \phi^\alpha
\end{eqnarray}
where we denote $\gamma_{n-2}=m_n$ and $\alpha_0=m_0$
while the gauge transformations (\ref{GT}) look as follows
\begin{eqnarray}\label{GTNorm}
\delta_0 \Psi^{\alpha(n)} &=& D \xi^{\alpha(n)} +
\frac{(\hat{s}+2)M}{2n(n+2)} e^\alpha{}_\beta
\xi^{\alpha(n-1)\beta} \nonumber \\
&& + \frac{2}{n(n-1)} m_n e^{\alpha\alpha} \xi^{\alpha(n-2)} + m_{n+2}
e_{\beta\beta} \xi^{\alpha(n)\beta\beta} \nonumber \\
\delta_0 \phi^\alpha &=& m_0 \xi^\alpha
\end{eqnarray}
Note that such gauge invariant description works not only in $AdS$
(and Minkowski) space but in $dS$ space as well provided
$m^2 > \frac{1}{4} \hat{s}^2\Lambda$, $\Lambda = - \lambda^2$ so that
massless limit is possible in the Minkowski and $AdS$ spaces only.
Inside the unitary forbidden region there exists a number of partially
massless cases \cite{DW01,DW01a,DW01c,Metsaev06.09,Zinoviev08.08}.
Namely each time when one of the parameters $m_n$ becomes zero the
whole system decomposes into two disconnected ones. One of them with
the fields $\Psi^{\alpha(2s-1)}\dots \Psi^{\alpha(n)}$ describes
partially massless field while the remaining fields describe massive
field with spin $\frac{n}{2}-1$.

\subsection{Formulation in terms of curvatures}

In this subsection we develop the formalism of gauge invariant
curvatures for massive higher spin fermionic field in $AdS_3$ and
rewrite the Lagrangian, found in previous subsection, in terms of
these curvatures. The gauge invariant curvatures have been
introduced before to construct the Lagrangian formulation for
massless higher spin fields in frame-like approach. We will show
that the objects with analogous properties can also be constructed
for massive higher spin fields. Since the gauge-invariant
description of massive fields uses the auxiliary Stueckelberg fields,
it is natural to construct the corresponding curvatures for them as
well.

First of all we introduce the additional auxiliary zero-forms
$C^{\alpha(n)},n=3,...,2(s-1)$\footnote{These fields are just the
first representatives of infinite number of zero-forms present in a
full unfolded formulation. They do not enter the free Lagrangian but
we need them to construct gauge invariant objects for physical
fields.} which transform under the gauge
transformations as follows
\begin{equation}\label{AddGT}
\delta_0 C^{\alpha(n)} = \eta_n \xi^{\alpha(n)}
\end{equation}
where the parameters $\eta_n$ are yet to be fixed. We combine these
fields with our spinor field denoting $\phi^\alpha=C^\alpha$ and
similarly for parameter of gauge transformations $\eta_1=\alpha_0$.
The most general ansatz for the full set of curvatures looks like
\begin{eqnarray*}
{\cal{R}}^{\alpha(n)} &=& D \Psi^{\alpha(n)} + \alpha_n
e^\alpha{}_\beta \Psi^{\alpha(n-1)\beta} + \beta_n e^{\alpha\alpha}
\Psi^{\alpha(n-2)} \\
&& + \gamma_n e_{\beta\beta} \Psi^{\alpha(n)\beta\beta} + f_n
E_2{}^\alpha{}_\beta {C}^{\alpha(n-1)\beta} \\
{\cal{C}}^{\alpha(n)} &=& D C^{\alpha(n)} + l_n \Psi^{\alpha(n)} + k_n
e^\alpha{}_\beta {C}^{\alpha(n-1)\beta} \\
&& + q_n e^{\alpha\alpha} C^{\alpha(n-2)} + p_n e_{\beta\beta}
C^{\alpha(n)\beta\beta}
\end{eqnarray*}
where $\alpha_n,\beta_n,\gamma_n$ are the same as in previous
subsection and guarantee that curvatures ${\cal{R}}$ is invariant
under the part of gauge transformation containing derivatives. Other
parameters are arbitrary and will be fixed by the gauge
invariance under the transformations (\ref{GT}),(\ref{AddGT}). Here
${\cal{C}}^{\alpha(n)}$ are the curvatures corresponding to the
zero-forms $C^{\alpha(n)}$.

Parameters $l_n,k_n,q_n,p_n$ are found from gauge invariance for
curvatures ${\cal{C}}$. We have
\begin{eqnarray*}
\delta{\cal{C}}^{\alpha(n)} &=& D (\eta_n \xi^{\alpha(n)}) + l_n
(D\xi^{\alpha(n)} + \alpha_n e^\alpha{}_\beta \xi^{\alpha(n-1)\beta} +
\beta_n e^{\alpha\alpha} \xi^{\alpha(n-2)} + \gamma_n e_{\beta\beta}
\xi^{\alpha(n)\beta\beta}) \\
&& + k_n e^\alpha{}_\beta (\eta_n \xi^{\alpha(n-1)\beta})
+ q_n e^{\alpha\alpha} (\eta_{n-2} \xi^{\alpha(n-2)})
+ p_n e_{\beta\beta} (\eta_{n+2} \xi^{\alpha(n)\beta\beta})
\end{eqnarray*}
It leads to the relations
\begin{align*}
& l_n + \eta_n = 0 && l_n\beta_n + q_n\eta_{n-2} = 0 \\
& l_n\alpha_n + k_n\eta_n = 0 && l_n\gamma_n + p_n\eta_{n+2} = 0
\end{align*}
Using the arbitrariness in definition of $C^{\alpha(n)}$ related
with multiplication by constant, $C^{\alpha(n)}\rightarrow
c_nC^{\alpha(n)}$ ($c_n$ are some numerical coefficients) we can set
$p_n=\gamma_n$. Then the general solution of the relations above looks
\begin{equation}\label{CurvParamC}
l_n = - \eta_n, \qquad k_n = \alpha_n, \qquad q_n = \beta_n, \qquad
\eta_n = \eta_{n+2}
\end{equation}
The last condition is the recurrent relation. To solve it we use as
initial data $\eta_1=\alpha_0$ and then obtain
$\eta_n = \eta_1 = \alpha_0$.

To find the last parameters $f_n$ we consider the invariance of the
curvature ${\cal{R}}^{\alpha(n)}$ under the gauge transformations
without derivatives. Invariance under the transformations containing
the derivatives has been established earlier. Corresponding variations
can be written in the form
\begin{eqnarray*}
\delta{\cal{R}}^{\alpha(n)} &=& - \lambda^2 E_2{}^\alpha{}_\beta
\xi^{\alpha(n-1)\beta} + \alpha_n( 4\alpha_n E_2{}^\alpha{}_\gamma
\xi^{\alpha(n-1)\gamma} + 2(n-2)\beta_n E_2{}^{\alpha\alpha}
\xi^{\alpha(n-2)} \\
&& + 8\beta_n E_2{}^{\alpha\alpha} \xi^{\alpha(n-1)} - 2n\gamma_n
E_2{}_{\gamma\gamma} \xi^{\alpha(n)\gamma\gamma}) +
\beta_n (-2(n-2)\alpha_{n-2} E_2{}^{\alpha\alpha} \xi^{\alpha(n-2)}
\\
&& - 2(n-1) \gamma_{n-2} E_2{}^\alpha{}_\beta \xi^{\alpha(n-1)\beta})
+ \gamma_n (2n\alpha_{n+2} E_2{}_{\beta\beta}
\xi^{\alpha(n)\beta\beta}  \\
&& + 2(n-1)\beta_{n+2} E_2{}_\beta{}^\alpha \xi^{\alpha(n-1)\beta}
+ 8\beta_{n+2} E_2{}_\beta{}^\alpha \xi^{\alpha(n-1)\beta})
+ f_n\eta_n E_2{}^\alpha{}_\beta \xi^{\alpha(n-1)\beta}
\end{eqnarray*}
Invariance condition leads to
\begin{align*}
& - \lambda^2 + 4\alpha_n{}^2 - 2(n-1)\beta_n\gamma_{n-2} +
2(n+3)\gamma_n\beta_{n+2} + f_n\eta_n = 0 \\
& 2(n+2)\alpha_n\beta_n - 2(n-2)\beta_n\alpha_{n-2} = 0
\end{align*}
Comparing with relations (\ref{Cond}), we conclude that
$$
f_1 = \alpha_0, \qquad f_n = 0, \qquad n \geq 3
$$
As a result we have the final expressions for the curvatures
\begin{eqnarray}\label{Curvatures}
{\cal{R}}^{\alpha(n)} &=& D \Psi^{\alpha(n)} + \alpha_n
e^\alpha{}_\beta \Psi^{\alpha(n-1)\beta} + \beta_n e^{\alpha\alpha}
\Psi^{\alpha(n-2)} + \gamma_n e_{\beta\beta}
\Psi^{\alpha(n)\beta\beta} \nonumber \\
{\cal{R}}^{\alpha} &=& D \Psi^{\alpha} + \alpha_1 e^\alpha{}_\beta
\Psi^{\beta} + \gamma_1 e_{\beta\beta} \Psi^{\alpha\beta\beta} +
\alpha_0 E_2{}^\alpha{}_\beta \phi^{\beta} \nonumber \\
{\cal{C}}^{\alpha} &=& D \phi^{\alpha} - \alpha_0 \Psi^{\alpha} +
\alpha_1 e^\alpha{}_\beta \phi^{\beta} + \gamma_1 e_{\beta\beta}
C^{\alpha\beta\beta}  \\
{\cal{C}}^{\alpha(n)} &=& D C^{\alpha(n)} - \alpha_0 \Psi^{\alpha(n)}
+ \alpha_n e^\alpha{}_\beta {C}^{\alpha(n-1)\beta} + \beta_n
e^{\alpha\alpha} C^{\alpha(n-2)} + \gamma_n e_{\beta\beta}
C^{\alpha(n)\beta\beta} \nonumber
\end{eqnarray}
Here the curvatures corresponding to $n=1$ are written separately.

Now let us rewrite the Lagrangian (\ref{MVLagr}) in terms of
curvatures. The most general expression for it has the form
\begin{eqnarray*}
{\cal{L}}_0 &=& i\sum_{n=1}^{2(s-1)} A_n
\Psi_{\alpha(n)}{\cal{R}}^{\alpha(n)} + i\sum_{n=1}^{2(s-1)} B_n
C_{\alpha(n-1)\beta} E_2{}^\beta{}_\gamma
{\cal{C}}^{\alpha(n-1)\gamma}
\end{eqnarray*}
where $A_n$ and $B_n$ are the arbitrary coefficients and are fixed
by requirement to reproduce initial Lagrangian (\ref{MVLagr}). Since
the fields $C^{\alpha(n)}$ for $n\geq3$ do not enter in
(\ref{MVLagr})  we can immediately put $B_n=0,n\geq3$. Redefining
$C^\alpha=\phi^\alpha,\;B_1=A_0$ ones get
\begin{eqnarray}\label{MVLagrCurv}
{\cal{L}}_0&=&i\sum_{n=1}^{2(s-1)}A_n\Psi_{\alpha(n)}{\cal{R}}^{\alpha(n)}+
iA_0\phi_{\alpha}E_2{}^\alpha{}_\beta{\cal{C}}^{\beta}
\end{eqnarray}
Comparing (\ref{MVLagrCurv}) with Lagrangian (\ref{MVLagr}) we
obtain
$$
A_n = \frac{\kappa_n}{2}, \qquad A_0 = \frac{1}{2}
$$
As a result we rewrite the Lagrangian (\ref{MVLagr}) in terms of
gauge invariant curvatures (\ref{MVLagrCurv}).

\subsection{Massive spin 5/2 example}

Massive field with spin $s=5/2$ is described by a set of fields
$\Psi^{\alpha(3)},\Psi^\alpha,\phi^{\alpha},C^{\alpha(3)}$, where the
second and third fields are Stueckelberg ones while the latter is an
auxiliary field and enter in the expressions for curvatures only. Then
in accordance with (\ref{MVLagrNorm}) the Lagrangian looks like
\begin{eqnarray*}
{\cal{L}}_0 &=& \frac{i}{2} \Psi_{\alpha(3)} D \Psi^{\alpha(3)} -
\frac{i}{2} \Psi_{\alpha} D \Psi^{\alpha} + \frac{i}{2} \phi_\alpha
E_2{}^\alpha{}_\beta D \phi^\beta \\
&& + im_3 \Psi_{\alpha(3)} e^{\alpha\alpha} \Psi^{\alpha} +
im_0 \Psi_{\alpha(1)} E_2{}^\alpha{}_\beta \phi^\beta \\
&& + \frac{iM}{2} \Psi_{\alpha(3)} e^\alpha{}_\beta
\Psi^{\alpha(2)\beta} - \frac{5iM}{6} \Psi_{\alpha} e^\alpha{}_\beta
\Psi^{\beta} - \frac{5iM}{2} \phi_\alpha E_3 \phi^\alpha
\end{eqnarray*}
where
$$
M^2 = m^2 + \frac94 \lambda^2, \qquad m{}^2 = 3m_3{}^2, \qquad
m_0{}^2 = 12(m^2 + 2\lambda^2)
$$
According to (\ref{Curvatures}) the expressions for the linearized
gauge-invariant curvatures have the form
\begin{eqnarray*}
{\cal{R}}^{\alpha(3)} &=& D \Psi^{\alpha(3)} + \frac{M}{3}
e^\alpha{}_\beta \Psi^{\alpha(2)\beta} +\frac{m_3}{3} e^{\alpha\alpha}
\Psi^{\alpha} \\
{\cal{R}}^{\alpha} &=& D \Psi^{\alpha} + \frac{5M}{3} e^\alpha{}_\beta
\Psi^{\beta} + m_3 e_{\beta\beta} \Psi^{\alpha\beta\beta} + m_0
E_2{}^\alpha{}_\beta \phi^{\beta} \\
{\cal{C}}^{\alpha} &=& D \phi^{\alpha} - m_0 \Psi^{\alpha} +
\frac{5M}{3} e^\alpha{}_\beta \phi^{\beta} + m_3 e_{\beta\beta}
C^{\alpha\beta\beta} \\
{\cal{C}}^{\alpha(3)} &=& D C^{\alpha(3)} - m_0 \Psi^{\alpha(3)} +
\frac{M}{3} e^\alpha{}_\beta{C}^{\alpha(2)\beta}
+ \frac{m_3}{3} e^{\alpha\alpha} \phi^{\alpha}
\end{eqnarray*}
The set of gauge transformations is written as follows
\begin{eqnarray*}
\delta_0 \Psi^{\alpha(3)} &=& D \xi^{\alpha(3)} + \frac{M}{3}
e^\alpha{}_\beta \xi^{\alpha(2)\beta} + \frac{m_3}{3} e^{\alpha\alpha}
\xi^{\alpha} \\
\delta_0 \Psi^{\alpha} &=& D\xi^{\alpha} + \frac{5M}{3}
e^\alpha{}_\beta \xi^{\beta} + m_3 e_{\beta\beta}
\xi^{\alpha\beta\beta} \\
\delta_0 \phi^\alpha &=& m_0 \xi^\alpha \\
\delta_0 C^{\alpha(3)} &=& m_0 \xi^{\alpha(3)}
\end{eqnarray*}
Accordingly to (\ref{MVLagrCurv}) the Lagrangian in term of
curvatures can be written in the form
\begin{eqnarray*}
{\cal{L}}_0 &=& \frac{i}{2} \Psi_{\alpha(3)} {\cal{R}}^{\alpha(3)} -
\frac{i}{2} \Psi_{\alpha} {\cal{R}}^{\alpha} + \frac{i}{2}
\phi_{\alpha} E_2{}^\alpha{}_\beta {\cal{C}}^{\beta}
\end{eqnarray*}

\section{Conclusion}

In this paper we have formulated a gauge-invariant Lagrangian
description for massive fermionic higher spins in three-dimensional
$AdS_3$ space. Using suitable set of Stueckelberg fields we have
derived the gauge-invariant Lagrangian and obtained a full set of
corresponding gauge transformations. We have also constructed a
complete set of linearized gauge-invariant curvatures on $AdS_3$ and
found that the gauge invariance required to introduce an additional
set of auxiliary fields which do not enter the free Lagrangian.
Massive spin-5/2 example is considered in details. We hope that our
results provides a ground for the development of frame-like
gauge-invariant formalism for interacting massive half-integer spins
in three dimensional space.

\appendix
\section*{Appendix}

In three dimensional space it is convenient to use two-component
spinor formalism and, in contrast to the $d=4$, only one type of
spinor indices is used. For instance $AdS_3$ background vielbein is
described by one-form $e^{\alpha(2)}$ and massless arbitrary
half-integer spin $s$ is described by one-form
$\Psi^{\alpha(n)},\;n=2(s-1)\;\mbox{is odd}$ (here $\alpha=1,2$ is
spinor index) where the argument of index means the number of
totally symmetrized indices. Further we present the list of our
notations and conventions for two-component spinor formalism and
differential form language (sign $\wedge$ of wedge product in the
text of the paper is everywhere omitted)
\begin{itemize}
\item For spinor indices labeled by one letter and standing on the
same level we use agreement of total symmetrization without
normalization factor for example
$$
e^\alpha{}_\beta \wedge \Psi^{\alpha(n-1)\beta} =
e^{(\alpha_1}{}_\beta \wedge \Psi^{\alpha_2...\alpha_n)\beta} =
e^{\alpha_1}{}_\beta \wedge
\Psi^{\alpha_2...\alpha_n\beta}+(n-1)\;\mbox{sym. terms}
$$
\item For anti-symmetric matrices
$\varepsilon^{\alpha\beta},\varepsilon_{\alpha\beta}$
we use the following basic relation and rule for lowering and
raising spinor indices
$$
\varepsilon^{\alpha\gamma} \varepsilon_{\gamma\beta} = -
\delta^\alpha{}_\beta, \qquad
\varepsilon^{\alpha\beta} A_\beta = A^\alpha, \qquad
\varepsilon_{\alpha\beta} A^\beta = - A_\alpha
$$
\item Basis elements of $1,2,3$-form spaces are respectively
$e^{\alpha(2)}$, $E_2{}^{\alpha(2)}$, $E_3$ where the last two are
defined as double and triple wedge product of $e^{\alpha(2)}$
$$
e^{\alpha\alpha} \wedge e^{\beta\beta} =
\varepsilon^{\alpha\beta}{E}_2{}^{\alpha\beta}
$$
$$
E_2{}^{\alpha\alpha} \wedge e^{\beta\beta} = \varepsilon^{\alpha\beta}
\varepsilon^{\alpha\beta} E_3
$$
Let us write useful relations for the basis elements
$$
E_2{}^\alpha{}_\gamma \wedge e^{\gamma\beta} = 3
\varepsilon^{\alpha\beta} E_3, \qquad
e^\alpha{}_\gamma \wedge e^{\gamma\beta} = 4 E_2{}^{\alpha\beta}
$$
$$
e^\alpha{}_\beta \wedge e^{\alpha\alpha} = 2\varepsilon_\beta{}^\alpha
E_2{}^{\alpha\alpha}, \qquad E_2{}^{\alpha}{}_\beta \wedge
e^{\alpha\alpha} = E_2{}^{\alpha\alpha} \wedge e^{\alpha}{}_\beta = 0
$$
\item For $AdS_3$ covariant derivative we use convention
$$
D \wedge D \xi^\alpha = - \lambda^2 E_2{}^{\alpha}{}_{\beta} \xi^\beta
$$
\end{itemize}

\section*{Acknowledgments}
I.L.B and T.V.S are grateful to the grant for LRSS, project No.
88.2014.2 and RFBR grant, project No. 12-02-00121-a for partial
support. Their research was also supported by grant of Russian
Ministry of Education and Science, project TSPU-122. T.V.S
acknowledges partial support from RFBR grant No. 14-02-31254. Work
of Yu.M.Z was supported in parts by RFBR grant No. 14-02-01172.


\newpage
\begin{thebibliography}{10}

\bibitem{Vasiliev96.11}
M.~A. Vasiliev.
{\it "Higher-spin gauge theories in four, three and two
dimensions",}
International J. Mod. Phys. {\bf D 05} (1996) 763.

\bibitem{BekCnockIazeolVasil05.03}
X.~Bekaert, S.~Cnockaert, C.~Iazeolla, and M.A. Vasiliev.
{\it "Nonlinear higher spin theories in various dimensions",}
arXiv:hep-th/0503128.

\bibitem{Vasiliev04.01}
M.~A. Vasiliev.
{it "Higher spin gauge theories in various dimensions",}
Fortschr. Phys. {\bf 52} (2005) 702, arXiv:hep-th/0401177.

\bibitem{DidenSkvorts14.01}
V.~E. Didenko and E.~D. Skvortsov.
{\it "Elements of {Vasiliev} theory",}
arXiv:1401.2975.

\bibitem{Metsaev06.09}
R.~R. Metsaev.
{\it "Gauge invariant formulation of massive totally symmetric
fermionic fields in {(A)dS} space",}
Phys. Lett. {\bf B643} (2006) 205, arXiv:hep-th/0609029.

\bibitem{Zinoviev08.08}
Yu.~M. Zinoviev.
{\it  "Frame-like gauge invariant formulation for massive high spin
particles",}
Nucl. Phys. {\bf B808} (2009) 185, arXiv:0808.1778.

\bibitem{PonomVasil10.01}
D.~S. Ponomarev and M.~A. Vasiliev.
{\it "Frame-like action and unfolded formulation for massive
higher-spin fields",}
Nucl. Phys. {\bf B839} (2010) 466, arXiv:1001.0062.

\bibitem{BuchKrykh05.05}
I.~L. Buchbinder and V.~A. Krykhtin.
{\it "Gauge invariant {Lagrangian} construction for massive
bosonic higher spin fields in {D} dimensions",}
Nucl. Phys. {\bf B727} (2005) 537, arXiv:hep-th/0505092.

\bibitem{BuchKrykhRyskTak06.03}
I.~L. Buchbinder, V.~A. Krykhtin, L.~L. Ryskina, and H. Takata.
{\it "Gauge invariant {Lagrangian} construction for massive higher
spin fermionic fields",}
Phys. Lett. {\bf B641} (2006) 386, arXiv:hep-th/0603212.

\bibitem{AragDeser84}
C. Aragone and S. Deser.
{\it "Hypersymmetry in D=3 of coupled gravity-massless spin-5/2
system",}.
Class. Quant. Grav. {\bf 1} 91984) L9.

\bibitem{Blencowe89}
M.~P. Blencowe.
{\it "A consistent interacting massless higher-spin field theory
in {D=2+1}",}
Class. Quant. Grav. {\bf 6} (1989) 443.

\bibitem{CamFredPfenTheis10.08}
A. Campoleoni, S. Fredenhagen, S. Pfenninger, and S. Theisen.
{\it "Asymptotic symmetries of three-dimensional gravity coupled
to higher-spin fields",}
JHEP {\bf 11} (2010) 007, arXiv:1008.4744.

\bibitem{ChenLongWang12.09}
B. Chen, J. Long, and Y. Wang.
{\it "Black holes in truncated higher spin $ads_3$ gravity",}
JHEP {\bf 12} (2012) 052, arXiv:1209.6185.

\bibitem{BergshHolmTowns09.11}
E.~A. Bergshoeff, O. Hohm, and P.~K. Townsend.
{\it "On higher derivatives in 3d gravities and higher spin gauge
theories",}
Annals Phys. {\bf 325} (2010) 1118, arXiv:0911.3061.

\bibitem{TyutVasil97.04}
I. Tyutin and M. Vasiliev.
{\it "Lagrangian formulation of irreducible massive fields of
arbitrary spin in 2+1 dimensions",}
Theor. Math. Phys. {\bf 113} (1997) 1244, arXiv:hep-th/9704132.

\bibitem{Zinoviev12.05}
Yu.~M. Zinoviev.
{\it "On massive gravity and bigravity in three dimensions",}
Class. Quant. Grav. {\bf 30} (2013) 055005, arXiv:1205.6892.

\bibitem{BuchSnegZin12.07}
I.~L. Buchbinder, T.~V. Snegirev, and Yu.~M. Zinoviev.
{\it "Gauge invariant lagrangian formulation of massive higher
spin fields in {$AdS_3$} space",}
Phys. Lett. {\bf B716} (2012) 243, arXiv:1207.1215.

\bibitem{BuchSnegZin12.08}
I.~L. Buchbinder, T.~V. Snegirev, and Yu.~M. Zinoviev.
{\it "On gravitational interactions for massive higher spins in
{$AdS_3$}",}
J. Phys. A {\bf 46} (2013) 214015, arXiv:.1208.0183

\bibitem{Vasiliev87}
M.~A. Vasiliev.
{\it "Free massless fields of arbitrary spin in the {de Sitter}
space and initial data for a higher spin superalgebra",}
Fortschr. der Physik {\bf 35} (1987) 741.

\bibitem{LV88}
V.~E. Loppatin and M.~A. Vasiliev.
{\it "Free massless bosonic fields of arbatrary spin in d-dimensional
de Sitter space",}
Mod. Phys. Lett. {\bf A3} (1988) 257.

\bibitem{Vas88}
M.~A. Vasiliev.
{\it "Free massless fermionoc fields of arbitrary spin in
d-dimensional de Sitter space",}
Nucl. Phys. {\bf B301} (1988) 26.

\bibitem{DW01}
S. Deser, A. Waldron
{\it "Gauge Invariance and Phases of Massive Higher Spins in (A)dS",}
Phys. Rev. Lett. {\bf 87} (2001) 031601, arXiv:hep-th/0102166.

\bibitem{DW01a}
S. Deser, A. Waldron
{\it "Partial Masslessness of Higher Spins in (A)dS",}
Nucl. Phys. {\bf B607} (2001) 577, arXiv:hep-th/0103198.

\bibitem{DW01c}
S. Deser, A. Waldron
{\it "Null Propagation of Partially Massless Higher Spins in (A)dS and
  Cosmological Constant Speculations",}
Phys. Lett. {\bf B513} (2001) 137, arXiv:hep-th/0105181.

\end{thebibliography}
\end{document}